\documentclass{elsart}
\usepackage{graphicx}
\begin{document}

\begin{frontmatter}

\title{Structure of isobaric analog states in $^{91}\mbox{Nb}$
populated by the $^{90}\mbox{Zr}(\alpha,\mbox{t})$ reaction} 

\author[kvi]{H.K.T. van der Molen},
\author[rcnp]{H. Akimune},
\author[kvi]{\underline{A.M. van den Berg}\thanksref{author}},
\author[rcnp]{I. Daito},
\author[rcnp]{H. Fujimura},
\author[osau]{Y. Fujita},
\author[rcnp,jaeri]{M. Fujiwara},
\author[kvi]{M.N. Harakeh},
\author[rcnp]{F. Ihara},
\author[rcnp]{T. Inomata},
\author[rcnp]{K. Ishibashi},
\author[umi]{J. J\"anecke},
\author[kvi]{N. Kalantar-Nayestanaki},
\author[ipno]{H. Laurent},
\author[ipno]{I. Lhenry},
\author[umi]{T.W. O'Donnell},
\author[kvi,mepi]{V.A. Rodin},
\author[kyou]{A. Tamii},
\author[spring8]{H. Toyokawa},
\author[kvi,mepi]{M.H. Urin},
\author[rcnp]{H. Yoshida} and
\author[rcnp]{M. Yosoi}
\address[kvi]{Kernfysisch Versneller Instituut, 
9747 AA Groningen, The Netherlands}
\address[rcnp]{Research Center for Nuclear Physics, Osaka University, 
Ibaraki, Osaka 567-0047, Japan}
\address[osau]{Department of Physics, Osaka University,
Toyonaka, Osaka 560-0043, Japan}
\address[jaeri]{Advanced Science Research Center, JAERI,
Tokai, Ibaraki, 319-1195, Japan}
\address[umi]{Department of Physics, University of Michigan,
Ann Arbor, Michigan 48109-1120, U.S.A.}
\address[ipno]{Institut de Physique Nucl\'eaire, IN2P3-CNRS, 
91406 Orsay Cedex, France}
\address[mepi]{Department of Theoretical Nuclear Physics, MEPI,
115409 Moscow, Russia}
\address[kyou]{Department of Physics, Kyoto University
Sakyo, Kyoto 606-8224, Japan}
\address[spring8]{SPring-8, Japan Synchrotron Research Institute,
Hyogo 679-5198, Japan}
\thanks[author]{Email: berg@kvi.nl}

\begin{abstract}
Decay via proton emission of isobaric analog states (IAS's) in
$^{91}\mbox{Nb}$ was studied using the $^{90}\mbox{Zr}(\alpha,t)$
reaction at $E_\alpha$=180 MeV. This study provides information about
the damping mechanism of these states. Decay to the ground state and
low-lying phonon states in $^{90}\mbox{Zr}$ was observed. The
experimental data are compared with theoretical predictions wherein
the IAS `single-particle' proton escape widths are calculated in a
continuum RPA approach. The branching ratios for decay to the phonon
states are explained using a simple model. 

\end{abstract}

\end{frontmatter}

\section{Introduction}

The study of the decay of single-particle (s.p.) or single-hole
states gives insight into their damping mechanisms. To expound on
this, consider the width of a s.p. state embedded in the continuum.
This is usually described by a width $\Gamma = \Gamma^\uparrow +
\Gamma^\downarrow$, where $\Gamma^\uparrow$ is the escape width and
$\Gamma^\downarrow$ is the spreading width. The escape width is
related to the direct decay by particle emission of a s.p. state.
This width is small for unbound states near the particle-decay
threshold, but increases rapidly as a function of excitation energy.
The spreading/fragmentation width is related to the coupling of the
state to more complex many-particle-many-hole states, where the
complexity increases up to statistical equilibrium. It is expected
that the first step in this damping mechanism is mediated via
coupling to a special class of particle-hole states, i.e. the surface
vibrations (phonon states) {\cite{gal88,van91}}. This mechanism plays
an important role at low excitation energies. From these phonon
states, acting as a doorway, coupling will take place to more complex
states until statistical equilibrium is reached. Valuable information
about the damping process and, therefore, on the structure of the
excited s.p. states can be obtained by the experimental determination
of the direct escape width and the width for semi-direct decay to
these phonon states. Since isobaric analogs of s.p. states have
similar structure as the s.p. states in the parent nucleus, the study
of their decay will reflect the damping mechanisms of the parent s.p.
state, i.e. their coupling to the continuum and its
spreading/fragmentation. 

In the seventies and eighties proton-stripping reactions on
medium-weight nuclei were used to study IAS's; see among others
Finkel {\it et al.} \cite{fin73,fin79}. Assume for simplicity that
the excess neutrons in a core nucleus $(Z,N)$ with $N>Z$ fill exactly
one complete neutron subshell with spin $j$. An extra neutron
(indicated with $\nu$) can be added in the next higher shell with
spin $j_\nu$. Then the IAS of this ground state (g.s.) has a simple
structure. If one applies the isospin-lowering operator to the g.s.
wave function, one gets two components: one part comes from the
transition of the valence neutron into the corresponding high-lying
proton orbit and the second part comes from the transition of a
neutron from the filled subshell into the corresponding open proton
orbit; thus forming a two-particle-one-hole state. Note, the second
part will only be possible for a core where $N>Z$. It should be
emphasized here that in the proton-stripping reaction on the core
nucleus $(Z,N)$ populating the IAS in the nucleus $(Z+1,N)$, only the
first part will contribute to the cross section. 

The proton decay of the analog of a single-neutron state of a
`core-plus-valence-neutron' in a parent nucleus can be described in
terms of the IAS s.p. proton decay populating the g.s. of the core
nucleus. Considerable efforts have been undertaken in the last three
decades to describe in a quantitative way the s.p. proton widths of
the IAS's (see e.g. Refs. \cite{aue83,gub86} and references therein).
However, only in the last decade the corresponding self-consistent
continuum Random-Phase Approximation (CRPA) or continuum Tamm-Dancoff
Approximation (CTDA) approaches have been developed and applied to
the closed-shell \cite{rum94,gia94,kno95} and to the
`closed-(neutron)shell + valence-neutron' parent nuclei \cite{rum94}.
In Ref. \cite{rum94}, the CRPA approach taken in the form developed
by Muraviev and Urin \cite{mur94} was used together with a
phenomenological mean field and the Landau-Migdal particle-hole
interaction. The partial self-consistency condition, which is a
result of the approximate isospin-symmetry conservation in nuclei
(see e.g. Ref. \cite{bir74}), was used by Rumyantsev and Urin
\cite{rum94} who explained a number of experimental data
satisfactorily. 

In the present work, the one-proton stripping $(\alpha,t)$ reaction
was studied to investigate the damping mechanism of IAS's in a
medium-weight nucleus. Because the $(\alpha,t)$ reaction is $Q$-value
mismatched, predominantly high-spin states were populated. First
results of this work have been reported elsewhere \cite{mol99}. In
addition to the population and decay of IAS's, data were also
obtained for high-lying s.p. states in the same experiment. Those
data and a full description of the experimental procedure and data
analysis will be published elsewhere \cite{mol**}. 

\section{Experimental procedure and results} 

The measurement of the $^{90}\mbox{Zr}(\alpha,t)$ proton-stripping
reaction was performed at the Research Center for Nuclear Physics
(RCNP). The beam energy used for the reaction was 180 MeV and the
triton ejectiles were momentum analyzed using the Grand Raiden
spectrometer \cite{fuj99}. The $^{90}\mbox{Zr}$ target used in the
experiment was a metallic foil with a thickness of 1 mg/cm$^2$ (98\%
enrichment). For calibration purposes, a natural carbon target was
used. Since the cross sections for exciting high-spin states in this
reaction peak at a scattering angle of $0^\circ$, the spectrometer
was set at $0.3^\circ$. The direct beam from the cyclotron was
stopped inside the spectrometer using a well-shielded beam dump. The
protons emitted from the excited states were detected in a 37-element
solid-state detector array, covering the backward hemisphere in the
angular range between $100^\circ$ and $160^\circ$. The
excitation-energy region covered by the setting of the spectrometer
ranged from 5 to 12 MeV in $^{91}\mbox{Nb}$. This region was chosen,
because it covers the range of known IAS's \cite{fin73,fin79} and
because the threshold for proton emission starts at an excitation
energy of 5.16 MeV. A singles and a coincidence spectrum are shown in
Fig. 1. In addition to some peaks originating from oxygen and carbon
contaminants in the target, sharp peaks are seen in the singles and
coincidence triton spectra. Comparing the singles and coincidence
spectra, it is clear that the large physical background observed in
the singles spectrum is reduced substantially because of the
detection of a coincident proton emitted in the backward direction,
i.e. the region where the breakup process yields very small cross
sections. Furthermore, it is clear from the coincidence spectrum,
that above the neutron-emission threshold at an excitation energy of
12.05 MeV, the yields for coincidences with protons drops
significantly. 

As the momenta for a triton ejectile and a proton emitted from a
$^{91}\mbox{Nb}$ nucleus were measured, it is possible to determine
the final-state energy $E_{fs} = E_x(^{91}\mbox{Nb}) - E_p - S_p -
E_r$, where $S_p$ and $E_r$ are the proton separation energy and the
energy of the recoiling $^{90}\mbox{Zr}$ nucleus, respectively. As an
example, Fig. \ref{plbfig2} shows the final-state energy spectrum for
the decay of states of $^{91}\mbox{Nb}$ in the excitation-energy
region between 11.85 and 12.20 MeV, where the $J^\pi = 11/2^-$ IAS is
located \cite{fin73}. Because the 37 proton detectors were mounted in
the backward hemisphere, in the angular range between $100^\circ$ and
$160^\circ$, it was possible to measure the angular correlation
between the direction of the protons emitted and the recoil axis of
the excited $^{91}\mbox{Nb}$ nuclei; see Fig. \ref{plbfig3} for the
decay of the IAS$(11/2^-)$. 

The analog state of the $J^\pi = 5/2^+$ $^{91}\mbox{Zr}$ g.s. is
located at $E_x = 9.86$ MeV in $^{91}\mbox{Nb}$ \cite{fin73}. Because
of the relatively small cross section, this IAS cannot be identified
in the present singles spectrum. However, in the coincidence
spectrum, this IAS is clearly seen; see Fig. 1. Our data are
consistent with earlier measurements done by Finkel {\it et al.}
\cite{fin79} who find that the decay of this state proceeds to the
g.s. of $^{90}\mbox{Zr}$ only. Since we cannot determine the singles
cross section for the population of this state, the absolute
branching ratio and, therefore, the escape width cannot be deduced
from the present data. For the higher lying IAS$(11/2^-)$, both the
singles cross section and the cross sections for decay to the lowest
states in $^{90}\mbox{Zr}$ have been extracted. Using the calculated
angular correlations shown in Fig. \ref{plbfig3}, the absolute
branching ratios $b_c$ for the decay by proton emission are obtained
by integrating them over $4\pi$ of the proton solid angle and
dividing this by the singles cross sections. The branching ratios are
listed in Table \ref{tab1} together with the values for the angular
momentum $l_c$ transferred in the decay channel. Since the width of
the IAS$(11/2^-)$ has not been determined experimentally, the partial
decay widths $\Gamma_c$ cannot be deduced from the present data. 

\section{Comparison with calculations} 

In the present work, we use the CRPA approach developed by Moukhai
{\it et al.} \cite{mou98} which has been applied recently
\cite{mur94} for the description of the IAS s.p. proton decay in
medium-heavy nuclei \cite{gor00}. This approach is equivalent to
earlier work \cite{rum94} but it is simpler in practice. In agreement
with Ref. \cite{gor00}, we find that the proton pairing can be
neglected in the description of the decay of the IAS's in
$^{91}\mbox{Nb}$ by proton emission, provided that the proton-decay
energy $\epsilon$ and the spectroscopic factor $S_\nu$ for the
valence-neutron states are taken from experiment\footnote{Note that
we do not use the convention $C^{2}S$ for the s.p. spectroscopic
factor, but rather $S_\nu$. For the IAS's of $^{91}\mbox{Nb}$ the
isospin Clebsch-Gordon coefficient is $C^{2} = 1/11$.}. This
assumption means that the calculated width $\Gamma_c^\mathrm{calc} =
S_\nu^\mathrm{exp} \cdot \Gamma_c^\mathrm{calc}(\mathrm{s.p.})$ with 
\begin{equation}
\label{e7}
\Gamma_c^\mathrm{calc}(\mathrm{s.p.}) = \Gamma_c^\mathrm{RPA}\cdot
\frac{1}{S_\nu}\cdot \frac{P_{l_\pi}(\epsilon^\mathrm{exp})}
{P_{l_\pi}(\epsilon)}
\end{equation}
can be compared with the corresponding experimental one. Here,
$c=\{l_\pi,j_\pi\}$ represents the quantum numbers of the emitted
proton with $l_\pi=l_\nu$ and $j_\pi=j_\nu$, and
$\Gamma_c^\mathrm{RPA}$ is the IAS s.p. width calculated within the
CRPA, taken at the calculated escape-energy $\epsilon$ of the proton.
In Eq. (\ref{e7}), $P_l(\epsilon)$ is the proton potential-barrier
penetrability (see e.g. Ref. \cite{boh69}) and $S_\nu^\mathrm{exp}$
is the experimental spectroscopic factor, while within the CRPA
$S_\nu=1$. In case of the $^{91}\mbox{Zr}$ parent nucleus we take
within the BCS model (see e.g. Ref. \cite{sol76}) the proton pairing
into account only to describe the proton separation energy. 

The results of the calculations given below have been obtained with
the use of the isoscalar part of the nuclear mean field taken from
Ref. \cite{mur94} and it contains four phenomenological parameters
(two strength parameters and two geometrical ones). The solution of
the BCS equations for the proton subsystem is found using only the
proton s.p. bound states. The intensity of the pairing forces for
protons, $G_p=27/A$ MeV, is chosen to reproduce the experimental
proton-pairing energy $P^{exp}(^{90}\mbox{Zr})=1.15$ MeV
\cite{wap85}. This intensity is close to the one mentioned by
Soloviev \cite{sol76}. The experimental neutron and proton separation
energies are reproduced in our calculations provided that $f'=0.96$
is chosen as the dimensionless amplitude of the isovector part of the
Landau-Migdal particle-hole interaction. This value of $f'$ is close
to the one used in earlier work \cite{rum94,mou98}. Thus, all model
parameters are fixed. In Table \ref{tab_sep_ener}, the calculated and
experimental separation energies are listed for $^{91}\mbox{Zr}$ and
$^{91}\mbox{Nb}$. 

Consider now the calculation of the parameters for the IAS($5/2^+)$,
which is the analog of the g.s. of $^{91}\mbox{Zr}$. The Fermi
strength function calculated for the $^{91}\mbox{Zr}$ parent nucleus
exhibits a narrow state with a peak energy $E_{IAS}=11.52$ MeV with
respect to the g.s. of $^{91}\mbox{Zr}$. The state corresponding to
the mentioned IAS$(5/2^+)$ in $^{91}\mbox{Nb}$ exhausts 96.5\% of the
Fermi sum rule $N-Z=11$. The calculated excitation energy of the
state $E_x = E_{IAS} - S_n (^{91}{\rm Zr}) + S_p(^{91}{\rm Nb})$ is
equal to 9.47 MeV. This value is in good agreement with the
experimental one, $E_x^{exp} = 9.86$ MeV, deduced from the
$^{90}\mbox{Zr}(^{3}\mbox{He},d)$ reaction \cite{fin73}. The s.p.
proton escape width of the IAS$(5/2^+)$ calculated with Eq.
(\ref{e7}) is equal to 3.3 keV. In this calculation we use the
spectroscopic factor $S_\nu^{exp} (d_{5/2}) = 0.95$ \cite{gra72} and
the experimental escape energy $\epsilon^{exp} = E_x^{exp}(^{91}{\rm
Nb}) - S_p^{exp}(^{91}{\rm Nb}) = 4.71$ MeV; ($\epsilon^{calc} =
4.32$ MeV, $\Gamma_c^{RPA} = 1.7$ keV). The calculated value agrees
well with the width $\Gamma^{exp}(2d_{5/2}) =4.0\pm0.5$ keV deduced
from the $^{90}\mbox{Zr}(p,p_{0})$ reaction \cite{sco69}. 

The present analysis of the s.p. proton width of the IAS($5/2^+)$ in
$^{91}\mbox{Nb}$ is refined as compared with the one given in Ref.
\cite{rum94}. In that work, the Coulomb mean field was calculated in
a non-consistent way and the spectroscopic factor $S_\nu = 0.86$ was
used. Nevertheless, the values of $\Gamma(2d_{5/2})$ calculated
presently and in Ref. \cite{rum94} differ little due to the partial
cancellation of the contributions to the width of the two mentioned
effects. 

The proton decay of the IAS$(5/2^+)$ to the $2^+$ phonon state in
$^{90}\mbox{Zr}$ is preferred as compared to the decay to the $5^-$
and $3^-$ states because the escaping proton has a higher energy and
the lowest angular momentum ($l_c=0$). The preliminary calculations
in an extended CRPA yield that the partial width for the decay to the
$2^+$ state has a value of several eV. This is in qualitative
agreement with the experimental result that the decay of the
IAS$(5/2^+)$ to the g.s. of $^{90}\mbox{Zr}$ exhausts about 100\% of
the total proton width of this IAS \cite{fin79}. 

The IAS$(11/2^-)$ has an excitation energy $E_x^{exp}=12.07$ MeV
\cite{fin73}. The excitation energy of the corresponding $11/2^-$
parent state in $^{91}\mbox{Zr}$ is 2.17 MeV which is very close to
the energy difference between the IAS$(11/2^-)$ and the IAS$(5/2^+)$
in $^{91}\mbox{Nb}$. The spectroscopic factor of this parent state is
$S_\nu^{(1)}=0.37$ ($S_\nu^{(1)}=0.45$) in Ref. \cite{gra72}
(\cite{boo69}). Another $11/2^-$ state in $^{91}\mbox{Zr}$ is located
at an excitation energy of 2.32 MeV and has a small spectroscopic
factor $S_\nu^{(2)}(11/2^-) = 0.05 $ \cite{gra72}. 

We propose simple wave functions to describe some of the measured
properties of the IAS$(11/2^-)$ state and the possible parent states
in $^{91}\mbox{Zr}$: 
\begin{equation}
\label{e14}
\Psi^{(i)}{(11/2^-)} = a_1^i \vert 1h_{11/2} \rangle +
a_2^i \vert 5^- \otimes 2d_{5/2} \rangle + a_3^i \vert 3^-
\otimes 2d_{5/2} \rangle , \,\,\, (i=1,2).
\end{equation}
The first term is due to the non-zero spectroscopic factors of the
$11/2^-$ parent states: i.e. $S_\nu^{(i)} = |a_1^i|^2$. In addition,
one has to account for the measured relative branching ratios for the
proton decay of the IAS$(11/2^-)$ to the $5^-_1$ and $3^-_1$ phonon
states in $^{90}\mbox{Zr}$ at $E_x = 2.32$ and 2.75 MeV,
respectively. For the parent states, this is achieved by coupling a
$d_{5/2}$ neutron to these negative-parity phonon states in
$^{90}\mbox{Zr}$ close in excitation energy to that of the $11/2^-$
states in $^{91}\mbox{Zr}$. Using the wave function given by Eq.
(\ref{e14}) and taking only the IAS s.p. proton decay into account,
we obtain the following expressions for the partial proton widths of
the IAS$(11/2^-)$ for the decay to the g.s., and to the $5^-$ and
$3^-$ states of the core nucleus: 
\begin{eqnarray}
\nonumber
\Gamma({\rm g.s.}) & = & |a_1|^2 \Gamma_{1}^{calc}(\mathrm{s.p.}), 
\,\,
1=1h_{11/2},\epsilon_1^\mathrm{exp},\\ \nonumber
\Gamma(5^-) & = &|a_2|^2 \Gamma_{2}^{calc} (\mathrm {s.p.}), \,\,
2=2d_{5/2},\epsilon_2^\mathrm{exp},\\
\Gamma(3^-)& = &|a_3|^2\Gamma_{3}^{calc}(\mathrm{s.p.}), \,\,
3= 2d_{5/2},\epsilon_3^\mathrm{exp},
\end{eqnarray}
where the subscript $c$ for the calculated s.p. width
$\Gamma_{c}^{calc} (\mathrm{s.p.})$ means that this width is
calculated using the indicated s.p. quantum numbers and the
experimental escape-energy $\epsilon^\mathrm{exp}$. The results of
the calculations are given in Table \ref{tab1}. The calculated s.p.
widths $\Gamma_{c}^{calc} (\mathrm{s.p.})$ are used to find the
experimental probabilities $|a_c|^2$ by comparing the relative
experimental branching ratios $\Gamma(c)/\Gamma({c'})$ and by using
the condition that $\sum_{c=1}^{3} |a_c|^2 = 1.0$. The resulting
values for $|a_c|^2$, which are obtained from the present experiment
are listed in Table \ref{tab1}. The value of the quantity
$|a_1|^2=0.45\pm0.03$ is in good agreement with the experimental
spectroscopic factor of the $11/2^-$ parent state $S_\nu(1h_{11/2}) =
0.37$ \cite{gra72}, $S_\nu(1h_{11/2}) =0.45$ \cite{boo69}. The
analysis of the experimental data on proton decay of the analog of
the rather complicated $11/2^-$ parent state in $^{91}\mbox{Zr}$ is
found to be reasonable. 

Theoretical evaluation of probabilities $|a_c|^2$ could be an
additional argument to judge the validity of the analysis. Such an
evaluation can be done in a coupled-channel approach using
phenomenological one-phonon transition potentials, which are
expressed via the dynamic-deformation parameter $\beta_L$ and the
nuclear mean field for neutrons $U^n(r)$ (see, e.g., Ref.
\cite{sam94}). Let $\nu_0$ and $\nu_1$ be the set of quantum numbers
for the $2d_{5/2}$ and $1h_{11/2}$ states, respectively; let $L^\pi$
and $\pi=(-1)^L$ be the phonon angular momentum and parity
$(L^\pi=5^-, 3^-)$. The coupling matrix elements are equal to: 
\begin{equation}
V^L_{\nu_1 \nu_0} 
= - \frac{\beta_L R}{\sqrt{2L+1}} 
\frac {\langle (\nu_1) || Y_L || (\nu_0) \rangle}
{\sqrt{2j_{\nu_1}+1}} \int 
\chi_{\nu_1}(r) \frac{\partial U^n}{\partial r} \chi_{\nu_0}(r)dr.
\end{equation}
Here, $R$ is the nuclear radius, $\langle (\nu_1) || Y_L || (\nu_0)
\rangle$ is the reduced matrix element and $r^{-1}\chi_{\nu}(r)$ are
the neutron bound-state radial wave functions. Using the model
parameters fixed previously and values for $\beta_L$ from Ref.
\cite{gra66} listed in Table \ref{tab1}, we find: $V^{L=5}_{\nu_1
\nu_0} = 0.37$ MeV and $V^{L=3}_{\nu_1 \nu_0} = 1.51$ MeV. The last
matrix element is comparable to the energy difference between the
$1h_{11/2}$ s.p. state ($E_x^{calc} = 4.40$ MeV) and the
configuration $(3^- \otimes 2d_{5/2})_{11/2^-} $ ($E_x^{exp} = 2.75$
MeV). This means that at least one of the two states of Eq.
(\ref{e14}) can have a rather large admixture of the $h_{11/2}$ s.p.
state (i.e. a rather large spectroscopic factor). In fact, we have a
three-level problem with known basis states having the mentioned
energies and with only two non-zero coupling matrix elements. The
solution of this problem gives three $11/2^-$ states with energies
$E_x^{calc} = 1.78$, 2.36 and 5.32 MeV. The calculated squared
amplitudes $|a_c|^2$ for the wave functions $\Psi^{(1,2)}$ of the two
states with the lowest excitation energies are given in Table
\ref{tab1}. The $11/2^-$ state at $E_x^{calc}=1.78$ MeV should be
assigned to the $11/2^-$ parent state at $E_x = 2.17$ MeV, because
the calculated and experimental spectroscopic factors for this state
are at least in qualitative agreement. For the same reason, the
second $11/2^-$ state at $E_x^{calc}=2.36$ should be assigned to the
$11/2^-$ state in $^{91}\mbox{Zr}$ at $E_x=2.32$ MeV. It should be
remarked that the above analysis of the structure of the $11/2^-$
states in $^{91}\mbox{Zr}$ at $E_x \sim$ 2 MeV can be considered only
as qualitative because rather schematic one-phonon transition
potentials have been used. A more detailed investigation can be made,
for instance, in the quasiparticle-phonon model \cite{sto00}.
Nevertheless, the performed analysis unravels the main structure of
these states. Increasing the number of basis states by taking into
consideration the $1g_{7/2}$ and $3s_{1/2}$ single-neutron states in
addition to the $2d_{5/2}$ state changes the calculated probabilities
$|a_c|^2$ and partial proton widths of the lowest two IAS$(11/2^-)$
only slightly. However, such an increase could lead to a strong
fragmentation of the strength of the third IAS$(11/2^-)$ at higher
excitation energies, which is presumably the reason why no third
$11/2^-$ state was observed in the experiment. 

\section{Conclusion} 

The one-neutron pickup spectroscopic factor $S_\nu$ of the $11/2^-$
parent state in $^{91}\mbox{Zr}$ found in the present analysis is in
agreement with the values found in Refs. \cite{gra72,boo69}. The
assumed structure of the $11/2^-$ parent state is qualitatively
confirmed by the calculations within the coupled-channel approach.
The abilities of the continuum-RPA approach are also checked by the
description of the partial proton width of the IAS($5/2^+$) in
$^{91}\mbox{Nb}$ for the decay to the g.s. of $^{90}\mbox{Zr}$. The
present analysis shows that the character of the IAS($5/2^+$) and its
parent state, i.e. the g.s. of $^{91}\mbox{Zr}$, is a pure $2d_{5/2}$
s.p. state coupled to the core nucleus $^{90}\mbox{Zr}$. The
IAS($11/2^-$) in $^{91}\mbox{Nb}$, however, is of a more complex
character; it consists of the combination of a $1h_{11/2}$ s.p. state
and a coupling of a $1d_{5/2}$ state to low-lying negative-parity
phonon states in $^{90}\mbox{Zr}$. 

\ack 

This work was performed as part of the research program of the {\it
Stichting voor Fundamentaal Onderzoek der Materie (FOM)} with
financial support from the {\it Nederlandse Organisatie voor
Wetenschappelijk Onderzoek (NWO)}. The authors (V.A.R. and M.H.U.)
acknowledge the support from {\it NWO} for their stay at the KVI. The
authors (A.M.v.d.B., M.N.H., N.K.-N. and H.K.T.v.d.M.) acknowledge
the support from the Japanese Ministry of Education ({\it Monbusho})
during their stay in Japan. 


\vfill
\begin{table}[h]
\caption{The weights $|a_c|^2$ of different components of the
$11/2^-$ parent state(s) in $^{91}\mbox{Zr}$ deduced from the
experimental data and calculated within the coupled-channel approach.}
\label{tab1}
\begin{center}
\begin{tabular}{|c|c|c|c|c|c|c|c|c|}
\hline
$c$     &$\epsilon_c^{exp}$ 
        &$l_c$
        &$b_c$
	&{$\Gamma_c^{calc}$(s.p.)$^{a)}$} 
	&\multicolumn{3}{c|}{$|a_c|^2$}	
        & $\beta_L$     \\ \cline{6-8} 
decay 	&	&	&	&	&experiment 	
&\multicolumn{2}{c|}{theory}	&\\ \cline{6-8}
channel &[MeV]  &[$\hbar$]      &          &[keV]  &        
&$\Psi^{(1)}$ 	&$\Psi^{(2)}$	&\\ \hline
g.s.    &6.90   &5      &$0.12\pm 0.01$ &0.42   &$0.45\pm 0.03$ 
&0.26 	&0.01	&--\\
$5^-$   &4.56   &0      &$0.27\pm 0.05$ &2.90   &$0.14\pm 0.03$ 
&0.12 	&0.88	&0.08\\
$3^-$   &4.16   &2      &$0.33\pm 0.03$ &1.25   &$0.41\pm 0.10$ 
&0.62 	&0.11	&0.20\\ \hline
\end{tabular}
\end{center}
$^{a)}$ Obtained from Eqs. (1) and (3).
\end{table}

\vfill
\begin{table}
\caption{Calculated and experimental separation energies.}
\label{tab_sep_ener}
\begin{tabular}{|c|c|c|c|c|}
\hline
nucleus	
&\multicolumn{2}{c|}{$S_n$ [MeV]}	
&\multicolumn{2}{c|}{$S_p$ [MeV]}\\
\cline{2-5}	&experiment &calculated &experiment &calculated \\ 
\hline
$^{91}\mbox{Zr}$&7.19	&7.11		&8.70	&8.55	\\
$^{91}\mbox{Nb}$&12.04  &12.04          &5.15   &5.15     \\      
\hline
\end{tabular}
\end{table}
\vfill

\begin{figure}
\begin{minipage}[htp]{75mm}
\includegraphics[width=75mm]{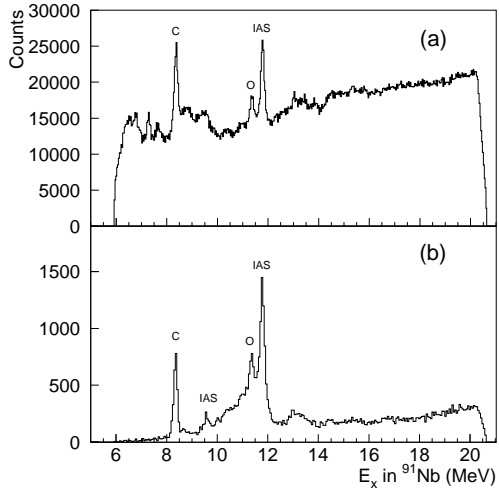}
\caption{
(a): The singles triton spectrum for the $^{90}\mbox{Zr}(\alpha,t)$
reaction at $E_\alpha = 180$ MeV and $\theta_t = 0.3^\circ$. (b): The
same spectrum as given in (a), but requiring in addition a
coincidence with a proton detected at backward angles. Indicated are
the IAS's and peaks due to contamination of the target with oxygen
(O) and carbon (C). 
}
\label{plbfig1}
\end{minipage}
\end{figure}

\begin{figure}
\begin{minipage}[htp]{75mm}
\includegraphics[width=75mm]{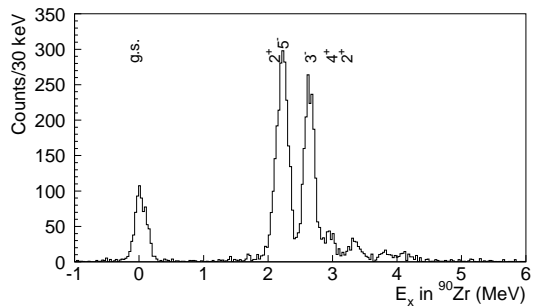}
\caption{Final-state spectrum after decay by proton emission 
from the excitation-energy region between 11.85 and 12.2 MeV. }
\label{plbfig2}
\end{minipage}
\end{figure}

\begin{figure}
\begin{minipage}[htp]{75mm}
\includegraphics[width=75mm]{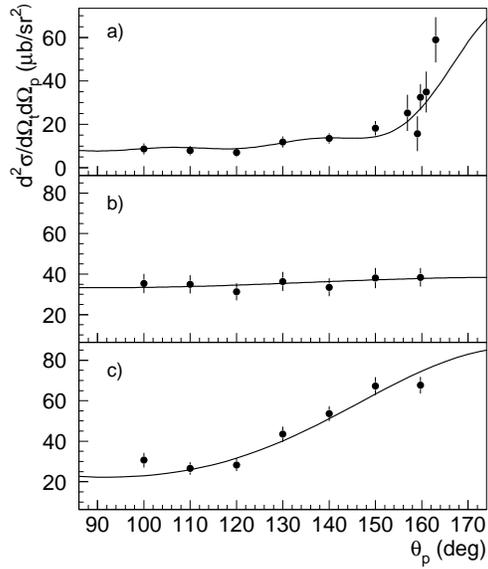}
\caption{Experimental (dots) and calculated (lines) angular
correlations for the decay from the 12 MeV IAS to different final
states in $^{90}$Zr; (a) to the g.s., (b) to the $2^+$/$5^-$ states
and (c) the $3^-$ state. The dominant $l$-values used for the
calculated angular correlations are listed in Table \ref{tab1}. Other
allowed $l$-values for decay to the $3^-$ and $5^-$ states are
included in the fit presented, but these have minor contributions and
no effect on the extracted branching ratio. 
} 
\label{plbfig3}
\end{minipage}
\end{figure}

\end{document}